\documentclass[aps,prb,reprint,twocolumn,superscriptaddress,floatfix,nofootinbib,longbibliography]{revtex4-2}
\usepackage{graphicx,amsmath,amsfonts,amssymb,amsthm,xr}
\usepackage{epsfig,amsmath,amssymb,color,dsfont,upgreek,physics}
\usepackage{mathrsfs}
\usepackage{mathtools}
\usepackage{bbold}
\usepackage{float}
\usepackage[caption=false]{subfig}
\usepackage[bookmarks=true,colorlinks,linkcolor=OrangeRed,urlcolor=NavyBlue,citecolor=RoyalBlue]{hyperref}
\usepackage[capitalize]{cleveref}
\usepackage[dvipsnames]{xcolor}
\usepackage{orcidlink}

\newcommand{\beginsupplement}{%
\pagebreak
\widetext
\setcounter{equation}{0}
\setcounter{figure}{0}
\setcounter{table}{0}
\setcounter{page}{1}
\makeatletter
\renewcommand{\theequation}{S\arabic{equation}}
\renewcommand{\thefigure}{S\arabic{figure}}
\renewcommand{\theHequation}{S\arabic{equation}}
\renewcommand{\theHfigure}{S\arabic{figure}}
\renewcommand{\thetable}{S\arabic{table}}
\renewcommand{\theHtable}{S\arabic{table}}
}%

\usepackage{tikz}
\usepackage{xstring}
\usepackage{preview}
\PreviewEnvironment{tikzpicture}

\newcommand{\circsize}{0.5ex}     
\newcommand{\circsep}{0.2em}      
\newcommand{\linewidthadjust}{0.08ex} 

\newcommand{\filledcircle}{%
  \tikz[baseline=-0.5ex] \fill (0,0) circle (\circsize);%
}
\newcommand{\emptycircle}{%
  \tikz[baseline=-0.5ex] \draw[line width=\linewidthadjust] (0,0) circle ({\circsize - 0.5*\linewidthadjust});%
}

\newcommand{\state}[1]{%
  \StrLen{#1}[\L]%
  \foreach \i in {1,...,\L}{%
    \StrChar{#1}{\i}[\c]%
    \ifnum\c=1
      \filledcircle
    \else
      \emptycircle
    \fi
    \ifnum\i<\L
      \hspace{\circsep}%
    \fi
  }%
}

\makeatother

\usepackage{babel}
\usepackage{orcidlink}
\newcommand{\Lmax}{36}
\begin{document}
\title{Non-Hermiticity-Enhanced Quantum Revivals in a Lattice Gauge Theory}

\author{Yevgeny Bar~Lev${}^{\orcidlink{0000-0002-1908-0771}}$}
\email{ybarlev@bgu.ac.il}
\affiliation{Department of Physics, Ben-Gurion University of the Negev, Beer-Sheva 84105, Israel}

\author{Jad C.~Halimeh${}^{\orcidlink{0000-0002-0659-7990}}$}
\email{jad.halimeh@lmu.de}
\affiliation{Department of Physics and Arnold Sommerfeld Center for Theoretical Physics (ASC), Ludwig Maximilian University of Munich, 80333 Munich, Germany}
\affiliation{Max Planck Institute of Quantum Optics, 85748 Garching, Germany}
\affiliation{Munich Center for Quantum Science and Technology (MCQST), 80799 Munich, Germany}
\affiliation{Department of Physics, College of Science, Kyung Hee University, Seoul 02447, Republic of Korea}

\author{Achilleas Lazarides${}^{\orcidlink{0000-0002-0698-2776}}$}
\email{a.lazarides@lboro.ac.uk}
\affiliation{Loughborough University, Loughborough, Leicestershire LE11 3TU, UK}

\begin{abstract}
  With the advent of quantum simulation experiments of lattice gauge theories (LGTs), the effect of non-Hermiticity on their physics has become an important question. The PXP model, a U$(1)$ LGT with a two-level electric field in one spatial dimension, is a paradigm of exotic physics out of equilibrium. Here, we consider a non-Hermitian version in which the spin-flip amplitude differs between the two spin directions. While the naive expectation is that non-Hermiticity might suppress coherent phenomena such as quantum many-body scars, when flips to the up (non-facilitating) state are favored, the oscillations are instead \emph{enhanced}. We show that this model can be mapped \emph{exactly} onto the standard PXP model by an imaginary gauge transformation for any value of the non-Hermiticity, with the transformation becoming singular only at the exceptional point (beyond which $g$ is complex); a longitudinal field maps onto a rescaled field in PXP. The mapping relates the enhanced revivals to the unchanged PXP dynamics and explains earlier numerical observations: From an initial N\'{e}el state, and because the non-Hermitian bias exponentially suppresses every non-N\'{e}el state while leaving the N\'{e}el amplitudes untouched, the first revival returns to the initial state with a fidelity per site that approaches 1, the deficit shrinking as $e^{-2g}$.  We discuss the implementation of this model as the no-click trajectory of a Lindbladian with local jump operators.
\end{abstract}

\maketitle
\global\long\def\E{\mathrm{e}}%
\global\long\def\D{\mathrm{d}}%
\global\long\def\I{\mathrm{i}}%
\global\long\def\mat#1{\mathsf{#1}}%
\global\long\def\vec#1{\mathsf{#1}}%
\global\long\def\cf{\textit{cf.}}%
\global\long\def\ie{\textit{i.e.}}%
\global\long\def\eg{\textit{e.g.}}%
\global\long\def\vs{\textit{vs.}}%

\global\long\def\ket#1{\left|#1\right\rangle }%

\global\long\def\etal{\textit{et al.}}%
\global\long\def\tr{\text{Tr}\,}%

\global\long\def\im{\text{Im}\,}%

\global\long\def\re{\text{Re}\,}%

\global\long\def\bra#1{\left\langle #1\right|}%

\global\long\def\braket#1#2{\left.\left\langle #1\right|#2\right\rangle }%

\global\long\def\obracket#1#2#3{\left\langle #1\right|#2\left|#3\right\rangle }%

\global\long\def\proj#1#2{\left.\left.\left|#1\right\rangle \right\langle #2\right|}%

\emph{Introduction.} Many-body quantum systems are, in general, ergodic: they thermalize, dynamically approaching a state well-described by statistical mechanics, as explained by the Eigenstate Thermalization Hypothesis (ETH)~\cite{Srednicki:1999bo,Deutsch:1991ju,Rigol:2008bf}, which states that eigenstates are themselves thermal, that is, locally indistinguishable from a thermal state at the same energy.

Lying at the heart of modern physics, gauge theories describe interactions between elementary particles mediated through gauge bosons \cite{Weinberg1995QuantumTheoryFields,Zee2003QuantumFieldTheory}. Their lattice variants, LGTs \cite{Gattringer2009QuantumChromodynamicsLattice,Rothe2012LatticeGaugeTheories}, were originally conceived to understand quark confinement \cite{Wilson1974ConfinementQuarks, Wilson1977QuarksStringsLattice}, but have since emerged as powerful tools beyond high-energy physics, lending deep insights into the nature of thermalization or lack thereof in quantum many-body systems \cite{Halimeh2025QuantumSimulationOutofequilibrium}.

In LGTs, violations of the ETH naturally arise from local constraints that restrict the accessible Hilbert space. Constrained LGTs have been shown to host a plethora of rich non-ergodic many-body phenomena, among which scarring is prominent~\cite{Surace2020LatticeGaugeTheories,Aramthottil2022ScarStates,Desaules2023WeakErgodicityBreaking,Desaules2023ProminentQuantumManyBodyScars,Osborne2024QuantumManyBodyScarring,Budde2024QuantumManyBodyScars}; see Refs.~\cite{Serbyn2021QuantumManyBodyScars,Moudgalya2022QuantumManyBodyScarsHilbertSpaceFragmentation,Chandran2023QuantumManyBodyScars} for reviews. Recent work towards quantum-simulating their physics \cite{Byrnes2006SimulatingLatticeGauge, Dalmonte2016LatticeGaugeTheory, Zohar2015QuantumSimulationsLattice, Aidelsburger2021ColdAtomsMeet, Zohar2021QuantumSimulationLattice, Klco2022StandardModelPhysics, Bauer2023QuantumSimulationHighEnergy, DiMeglio2024QuantumComputingHighEnergy, Halimeh2023ColdatomQuantumSimulators} has led to a flurry of experiments on state-of-the-art analog and digital quantum hardware, see, e.g., Refs.~\cite{Martinez2016RealtimeDynamicsLattice, Schweizer2019FloquetApproachZ2, Mil2020ScalableRealizationLocal, Yang2020ObservationGaugeInvariance, Su2023ObservationManybodyScarring, Zhou2022ThermalizationDynamicsGauge, Zhang2025ObservationMicroscopicConfinement, Gonzalez-Cuadra2025ObservationStringBreaking, De2024ObservationStringbreakingDynamics, Cochran2025VisualizingDynamicsCharges} and Ref.~\cite{Halimeh2025QuantumSimulationOutofequilibrium} for a review. This motivates further exploration of their physics beyond the conventional Hermitian setting that has dominated experiments so far.
\begin{figure}[t]
  \usetikzlibrary{calc}
  \begin{tikzpicture}[font=\normalsize]
    \def\xsep{1.4}
    \pgfmathsetmacro{\a}{0.5*\xsep}
    \def\sep{0.2cm}      

    \def\ktwoTop{1010}
    \def\kzero{0000}
    \def\ktwoBottom{0101}

    \node (T) at (0, 2) {$\left|\state{\ktwoTop}\right\rangle$};
    \node (Uleft)  at ({(-\a)}, 1) {$\left|\state{1000}\right\rangle$};
    \node (Uright) at ({(\a)},  1) {$\left|\state{0010}\right\rangle$};
    \node (M) at (0, 0) {$\left|\state{\kzero}\right\rangle$};
    \node (Dleft)  at ({(-\a)}, -1) {$\left|\state{0001}\right\rangle$};
    \node (Dright) at ({(\a)},  -1) {$\left|\state{0100}\right\rangle$};
    \node (B) at (0, -2) {$\left|\state{\ktwoBottom}\right\rangle$};

    \pgfmathsetmacro{\xcol}{1.5*\xsep}
    \node at (\xcol,  2.6) {$N_{\uparrow}$};
    \node at (\xcol,  2.0) {$2$};
    \node at (\xcol,  1.0) {$1$};
    \node at (\xcol,  0.0) {$0$};
    \node at (\xcol, -1.0) {$1$};
    \node at (\xcol, -2.0) {$2$};

    \pgfmathsetmacro{\xcol}{2*\xsep}
    \node at (\xcol,  2.6) {$V_{\uparrow}$};
    \node at (\xcol,  2.0) {$e^{2g}$};
    \node at (\xcol,  1.0) {$e^{g}$};
    \node at (\xcol,  0.0) {$1$};
    \node at (\xcol, -1.0) {$e^{g}$};
    \node at (\xcol, -2.0) {$e^{2g}$};


    \draw[->,red]  ([xshift=-\sep]M.north) -- ([xshift=-\sep]Uleft.south);
    \draw[->,blue]
    let \p1 = ([xshift= \sep]Uleft.south),
    \p2 = ([xshift= \sep]M.north),
    \p3 = ([xshift=-\sep]Uleft.south),
    \p4 = ($(\p1)!(\p3)!(\p2)$)
    in
    (\p4) -- ($(\p4) + (\p2) - (\p1)$);

    \draw[->,red]  ([xshift= \sep]M.north) -- ([xshift= \sep]Uright.south);
    \draw[->,blue]
    let \p1 = ([xshift=-\sep]Uright.south),
    \p2 = ([xshift=-\sep]M.north),
    \p3 = ([xshift= \sep]Uright.south),
    \p4 = ($(\p1)!(\p3)!(\p2)$)
    in
    (\p4) -- ($(\p4) + (\p2) - (\p1)$);

    \draw[->,red]  ([xshift=-\sep]M.south) -- ([xshift=-\sep]Dleft.north);
    \draw[->,blue]
    let \p1 = ([xshift= \sep]Dleft.north),
    \p2 = ([xshift= \sep]M.south),
    \p3 = ([xshift=-\sep]Dleft.north),
    \p4 = ($(\p1)!(\p3)!(\p2)$)
    in
    (\p4) -- ($(\p4) + (\p2) - (\p1)$);

    \draw[->,red]  ([xshift= \sep]M.south) -- ([xshift= \sep]Dright.north);
    \draw[->,blue]
    let \p1 = ([xshift=-\sep]Dright.north),
    \p2 = ([xshift=-\sep]M.south),
    \p3 = ([xshift= \sep]Dright.north),
    \p4 = ($(\p1)!(\p3)!(\p2)$)
    in
    (\p4) -- ($(\p4) + (\p2) - (\p1)$);

    \draw[->,red]  ([xshift=-\sep]Uleft.north) -- ([xshift=-\sep]T.south);
    \draw[->,blue]
    let \p1 = ([xshift= \sep]T.south),
    \p2 = ([xshift= \sep]Uleft.north),
    \p3 = ([xshift=-\sep]T.south),
    \p4 = ($(\p1)!(\p3)!(\p2)$)
    in
    (\p4) -- ($(\p4) + (\p2) - (\p1)$);

    \draw[->,red]  ([xshift= \sep]Uright.north) -- ([xshift= \sep]T.south);
    \draw[->,blue]
    let \p1 = ([xshift=-\sep]T.south),
    \p2 = ([xshift=-\sep]Uright.north),
    \p3 = ([xshift= \sep]T.south),
    \p4 = ($(\p1)!(\p3)!(\p2)$)
    in
    (\p4) -- ($(\p4) + (\p2) - (\p1)$);

    \draw[->,red]  ([xshift=-\sep]Dleft.south) -- ([xshift=-\sep]B.north);
    \draw[->,blue]
    let \p1 = ([xshift= \sep]B.north),
    \p2 = ([xshift= \sep]Dleft.south),
    \p3 = ([xshift=-\sep]B.north),
    \p4 = ($(\p1)!(\p3)!(\p2)$)
    in
    (\p4) -- ($(\p4) + (\p2) - (\p1)$);

    \draw[->,red]  ([xshift= \sep]Dright.south) -- ([xshift= \sep]B.north);
    \draw[->,blue]
    let \p1 = ([xshift=-\sep]B.north),
    \p2 = ([xshift=-\sep]Dright.south),
    \p3 = ([xshift= \sep]B.north),
    \p4 = ($(\p1)!(\p3)!(\p2)$)
    in
    (\p4) -- ($(\p4) + (\p2) - (\p1)$);

  \end{tikzpicture}

  \caption{\label{fig:hilbert-space}Hilbert subspace of $L=4$ system, which
    includes the N\'{e}el states, $\mathbb{Z}_{2}$ and $\bar{\mathbb{Z}}_{2}$.
    The columns indicate the number of up spins, $N_{\uparrow}$, and
    the corresponding weight of these states, $V_{\uparrow}$ (see Eq.~\eqref{eq:similarity_transformation}).
    The red (blue) arrows indicate transitions which raise (lower) $N_{\uparrow}$
    and have amplitude $e^{g}$ $\left(e^{-g}\right)$.}
\end{figure}
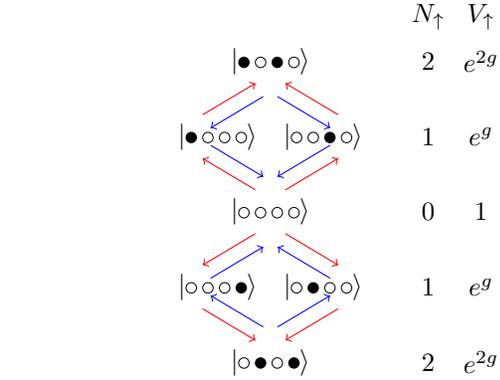

Ergodicity breaking can be strong or weak. In the strong case, exemplified by disordered systems that exhibit Many-Body Localization (MBL)~\cite{Gornyi2005InteractingElectrons,Basko2006MBL,Oganesyan2007MBL}, a finite fraction of the spectrum displays non-ergodic behavior. By contrast, in weak ergodicity breaking, of which the PXP model~\cite{LesanovskyPXP} with its scars~\cite{Bernien2017,Turner2018,TurnerPRB2018,Ho2019PeriodicOrbits,Choi2019EmergentSU2,Bluvstein2021Controlling} is probably the best-known example, a few states exhibit non-thermal behavior while all others thermalize.  In this model, there exist  $\mathcal{O}(L)$ (with $L$ the system size) non-thermal eigenstates, which display low entanglement entropy and are approximately equidistant in energy. Therefore, initial states that have significant overlap with them will display persistent oscillations in their return probability. The two N\'{e}el states have this property, so initializing the system in one of them results in the spins rotating coherently for a long time, despite the highly interacting nature of the model.

While long-lived, these oscillations do eventually decay in time. Their lifetime can be extended by Hermitian deformations of the model~\cite{Choi2019EmergentSU2,Bull2020QuantumScars}. Here, we instead consider a minimal non-Hermitian modification, in which spin flips up and down have unequal amplitudes. This turns out to sharpen the oscillations and enhance their lifetime. Non-Hermitian descriptions provide a minimal and intuitive description of open quantum systems~\cite{ashida_non-hermitian_2020} with loss, gain, or continuous monitoring~\cite{daley_quantum_2014}, and are now a large and growing field of research including first works in LGTs \cite{Cheng2024DynamicalLocalizationTransition,Hu2025ManyBodyNonHermitianSkinEffect}.

The key mechanism here is the following. The constraints bound the maximum magnetization by $N_{\uparrow}\le L/2$, with equality at the two N\'{e}el states. At the same time, our dynamics is related to that of the unitary PXP via a similarity transformation that, for $g>0$, reweights Fock states enhancing the amplitudes of states exponentially with $N_{\uparrow}$. As a result, all states with intermediate magnetization are exponentially (with g) suppressed, effectively enhancing the amplitude on the N\'{e}el states. This modifies the time evolution so as to concentrate weight onto the two N\'{e}el states. The mechanism is reminiscent of controlled bias in large deviation theory \cite{jack2010large} while our transformation is analogous to the imaginary gauge transformation of \cite{Hatano1996Localization}.

The Hamiltonian we study coincides, up to an overall factor, with the zero-field non-Hermitian PXP model of Ref.~\cite{Chen2023WeakErgodicityBreaking}, $\sum_j\hat{P}_{j-1}(\hat{\sigma}^x_j+i\gamma\hat{\sigma}^y_j)\hat{P}_{j+1}=\sqrt{1-\gamma^2}\,\hat{H}_g$ with $\gamma=\tanh g$. There, non-Hermitian scars were identified numerically, enhanced revivals with a period that diverges upon approaching the exceptional point $\gamma\to1$ were reported, and a perturbative embedding into a Lindbladian was proposed. The same model was considered in the Supplemental Material of Ref.~\cite{Shen2024EnhancedManyBody}, where no accumulation in the Hamming distance from $\mathbb{Z}_2$ (``Fock skin effect'') was found, while an enhancement of revivals for models with sublattice-dependent non-Hermiticity was explained within the forward-scattering approximation.

Our contribution is an exact similarity transformation, valid for every real $g$, which maps the model onto the Hermitian PXP model. It explains why the spectrum is real for $|\gamma|<1$, identifies the exceptional point with the singular limit $g\to\infty$ of the transformation, shows that the divergence of the period is the rescaling $T=T_{\mathrm{PXP}}/\sqrt{1-\gamma^2}$, extends to the longitudinal field considered in Ref.~\cite{Chen2023WeakErgodicityBreaking}, which maps onto a rescaled field $h_{\mathrm{eff}}=h/\sqrt{1-\gamma^{2}}$ (Sec.~S2 of~\cite{SM}), and gives the enhancement of the revivals in closed form in terms of PXP quantities. It also shows that our enhancement is an exactly mappable instance of the same broad mechanism as the Fock skin effect, namely a non-reciprocal bias in Fock space towards the scarred states, but with the bias in the magnetization, towards \emph{both} N\'{e}el states, which explains why it is invisible in the Hamming distance from one of them, and the same construction maps the models of Ref.~\cite{Shen2024EnhancedManyBody} onto Hermitian PXP chains with rescaled amplitudes~\cite{SM}. Revivals can also be enhanced by periodic projective measurements, even without postselection~\cite{Paviglianiti2025EnhancingRevivals}. Non-Hermitian versions of the PXP model have also been considered in the context of exceptional points and Yang-Lee criticality~\cite{Zhang2025NH_PXP_YangLee}, where the emphasis is on spectral singularities~\cite{bender_making_2007} rather than on scarred dynamics.

In what follows, we introduce our model and then numerically demonstrate that the oscillations are enhanced over a range of the control parameter. We next introduce the similarity transformation, and use it to analytically explain the enhancement of the oscillations. Finally, we study the spectral properties of our model and find that for $g>0$ the features that distinguish the scarred eigenstates from the rest in PXP are also enhanced.

\emph{Model and numerics.} We consider a non-Hermitian version of the PXP model,
\begin{equation}
  \hat{H}_{g}=\sum_{i=1}^{L}\hat{P}_{i-1}\left(e^{g}\hat{\sigma}_{i}^{+}+e^{-g}\hat{\sigma}_{i}^{-}\right)\hat{P}_{i+1},
\end{equation}
where $\hat{P}_{i}=\frac{1}{2}\left(1-\hat{Z}_{i}\right)$ is a projector
on the facilitating, $\downarrow$-state on site $i$, $\hat{Z}_{i}=\hat{\sigma}_{i}^{z}$,
the corresponding raising (lowering) operator is $\hat{\sigma}_{i}^{+}$
$\left(\hat{\sigma}_{i}^{-}\right)$, $L$ is the (even) number of sites
and $g$ is a real parameter. We use periodic boundary conditions
(PBC), namely, $\hat{P}_{0}=\hat{P}_{L}$ and $\hat{P}_{L+1}=\hat{P}_{1}$.
Following Ref.~\cite{Turner2018}, we designate the states
using \state{0} and \state{1}, which correspond to $\downarrow$
and $\uparrow$. For example, the N\'{e}el states will be given by $\mathbb{Z}_{2}\equiv\ket{\state{010101}}$,
and $\bar{\mathbb{Z}}_{2}\equiv\ket{\state{101010}}$. Hamiltonian dynamics does not affect two or more consecutive up spins and so the Hilbert space is composed of a large number of disconnected sub-spaces. Spatially, the dynamics to the left and right of an island of two (or more) spins is independent, such that the system is also spatially disconnected if one or more of such islands exist. In what follows, we will operate in the subspace of the Hilbert
space which contains the N\'{e}el states, $\mathbb{Z}_{2}$ and $\bar{\mathbb{Z}}_{2}$ (the Fibonacci or Rydberg-blockaded subspace),
see Fig.~\ref{fig:hilbert-space}, and in which there are no islands of consecutive up spins. In the language of LGTs, PXP is the spin-$1/2$ U$(1)$ quantum link model in which the matter fields have been integrated out using Gauss's law, which becomes the constraint~\cite{Surace2020LatticeGaugeTheories}. Each spin flip then corresponds to a gauge-invariant process, in which an electric field on a link flips and a particle-antiparticle pair is created or annihilated. Non-zero $g$ assigns different amplitudes to such a process and its Hermitian conjugate, while $\hat{H}_{g}$ still acts within the gauge-invariant subspace. In this mapping, the two N\'{e}el states correspond to the charge-free vacua, and each pair of adjacent down spins to a matter excitation, such that the number of matter excitations is $L-2N_{\uparrow}$; $g>0$ thus favors pair annihilation over pair creation, and the enhancement below is a reweighting towards the vacua.

For $g=0$, the model reduces to the standard Hermitian PXP model with
the Hamiltonian $\hat{H}_{0}$. This model is known to have $L+1$
special eigenstates, called quantum scars. These have atypically
low entanglement entropy and atypically high overlap with the N\'{e}el
states \cite{Turner2018}. Moreover, the states have an almost
constant separation in energy. Taking the N\'{e}el state as an initial
state, the system exhibits (non-perfect) oscillations between the
N\'{e}el states (top and bottom row in Fig.~\ref{fig:hilbert-space}),
with a period which is related to the energy separation \cite{Turner2018}.
For non-zero $g$, the amplitude to increase (decrease) the
number of up spins by one is weighted by $e^{g}$ $\left(e^{-g}\right)$.
Therefore, for $g>0$ the system flows towards the N\'{e}el states, while
for $g<0$ it flows towards the all-down state, $\ket{\state{0000}}$.
Ostensibly, for any value of $g$, the non-Hermiticity destroys oscillations
between the N\'{e}el states, since it biases against the transition between
the two N\'{e}el states. In what follows, we show that this is not true
for $g>0$.

\begin{figure}[!t]
  \includegraphics[width=1\columnwidth]{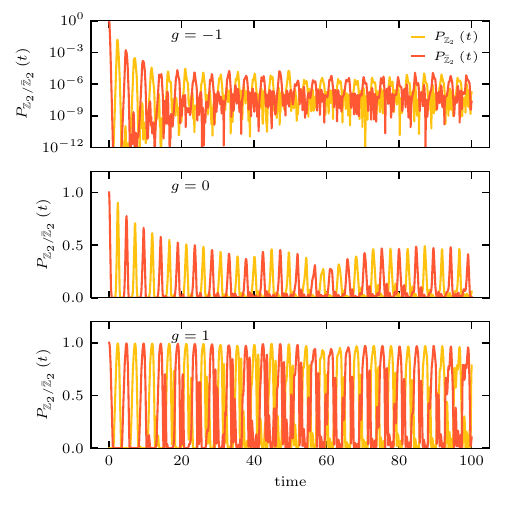}

  \caption{\label{fig:probability-z2-z2bar}Probability $P_{\mathbb{Z}_{2}/\bar{\mathbb{Z}}_{2}}\left(t\right)$
    of finding the system in $\mathbb{Z}_{2}$ or $\bar{\mathbb{Z}}{}_{2}$
    states as a function of time. The plots display results for $g=-1,0,1$
    and $L=18$ and initial state of $\bar{\mathbb{Z}}{}_{2}$; the top panel is on a logarithmic scale.}
\end{figure}

In Fig.~\ref{fig:probability-z2-z2bar}, we show the numerically
computed probability of finding the system in $\mathbb{Z}_{2}$ or $\bar{\mathbb{Z}}{}_{2}$
states, $P_{\mathbb{Z}_{2}}\left(t\right)=\left|\braket{\mathbb{Z}_{2}}{\bar{\mathbb{Z}}_{2}\left(t\right)}\right|^{2}/\braket{\bar{\mathbb{Z}}_{2}\left(t\right)}{\bar{\mathbb{Z}}_{2}\left(t\right)}$
and $P_{\bar{\mathbb{Z}}{}_{2}}\left(t\right)=\left|\braket{\bar{\mathbb{Z}}_{2}}{\bar{\mathbb{Z}}_{2}\left(t\right)}\right|^{2}/\braket{\bar{\mathbb{Z}}_{2}\left(t\right)}{\bar{\mathbb{Z}}_{2}\left(t\right)}$,
where $\ket{\bar{\mathbb{Z}}_{2}\left(t\right)}=\exp\left[-i\hat{H}_{g}t\right]\ket{\bar{\mathbb{Z}}_{2}}$.
The division by the norm of the states takes into account the
change in the norm of the state, which follows from the fact that
$\hat{H}_{g}$ is non-Hermitian for non-zero $g$~\cite{ashida_non-hermitian_2020}. We see that while the
amplitudes decay fast to zero for $g<0$, for $g>0$ they show sharp
oscillations between the N\'{e}el states, which do not appear to decay
strongly with time on the scale shown (for $g=1$ and $L=18$, all revival peaks up to $t=100$ exceed $0.93$), a regime which we call \emph{non-Hermiticity-enhanced quantum revivals}. In what follows, we provide an analytical explanation
of this behavior.

\emph{Similarity transformation and the exact enhancement bound.} A key numerical observation is that the spectrum of $\hat{H}_{g}$
does not depend on $g$, suggesting that the non-Hermitian $\hat{H}_{g}$
and the Hermitian $\hat{H}_{0},$ are related via a similarity
transformation,
\begin{equation}
  \hat{V}_{\uparrow}\hat{H}_{0}\hat{V}_{\uparrow}^{-1}=\hat{H}_{g}.\label{eq:similarity}
\end{equation}
It can be verified that the transformation,
\begin{equation}
  \hat{V}_{\uparrow}=\mathrm{e}^{g\hat{N}_{\uparrow}},\qquad\hat{N}_{\uparrow}=\frac{1}{2}\sum_{i=1}^{L}\left(1+\hat{Z}_{i}\right)\label{eq:similarity_transformation}
\end{equation}
has this property, since $\hat{V}_{\uparrow}\hat{\sigma}_{i}^{\pm}\hat{V}_{\uparrow}^{-1}=\mathrm{e}^{\pm g}\hat{\sigma}_{i}^{\pm}$, while the projectors $\hat{P}_{i}$ commute with $\hat{V}_{\uparrow}$. Here $\hat{N}_{\uparrow}$ is an operator that
counts the number of spins up. For real $g$, the transformation is Hermitian, but not unitary,
and diagonal in the computational basis. Eq.~\eqref{eq:similarity} holds on the full Hilbert space and for every real $g$; the spectrum of $\hat{H}_{g}$ is therefore real and identical to that of PXP, and the right eigenvectors of $\hat{H}_{g}$ become increasingly non-orthogonal as $|g|$ grows, the condition number of $\hat{V}_{\uparrow}$ being $\mathrm{e}^{|g|L/2}$ in the constrained subspace. In terms of the model of Ref.~\cite{Chen2023WeakErgodicityBreaking}, the exceptional point $\gamma\to1$ corresponds to $g\to\infty$, where $\hat{V}_{\uparrow}$ becomes singular. See Fig.~\ref{fig:hilbert-space} for an example of its operation on the basis states.

Using Eq.~\eqref{eq:similarity} and \eqref{eq:similarity_transformation}
we can obtain the evolution of any basis state, $\ket n$
\begin{align}
  \obracket m{\mathrm{e}^{-i\hat{H}_{g}t}}n & =\obracket m{\mathrm{e}^{g\hat{N}_{\uparrow}}\mathrm{e}^{-i\hat{H}_{0}t}\mathrm{e}^{-g\hat{N}_{\uparrow}}}n\nonumber                  \\
                                            & =\mathrm{e}^{g\left(N_{\uparrow}^{m}-N_{\uparrow}^{n}\right)}\obracket m{\mathrm{e}^{-i\hat{H}_{0}t}}n.\label{eq:generic_basis_state}
\end{align}
The evolution for any nonzero $g$ is, up to  normalization, the same as under the unitary evolution using the PXP Hamiltonian but with each product state weighted by the
exponential factor $\mathrm{e}^{g\left(N_{\uparrow}^{m}-N_{\uparrow}^{n}\right)}$.
For $g>0$ ($g<0$), this favors states with a larger (smaller) number
of up spins, compared to the initial state. Specifically, for the
N\'{e}el states, we see that $\braket{\mathbb{Z}_{2}}{\bar{\mathbb{Z}}_{2}\left(t\right)}$
and $\braket{\bar{\mathbb{Z}}_{2}}{\bar{\mathbb{Z}}_{2}\left(t\right)}$
do \emph{not} depend on $g$, and therefore the enhancement of the
revivals of the $\ket{\bar{\mathbb{Z}}_{2}}$ observed in Fig.~\ref{fig:probability-z2-z2bar} stems from suppression of the weights of the other states, as will be explained below.

Similarly to the derivation of Eq.~\eqref{eq:generic_basis_state},
we can compute the normalization,
\begin{align}
  \braket{\bar{\mathbb{Z}}_{2}\left(t\right)}{\bar{\mathbb{Z}}_{2}\left(t\right)} & =\mathrm{e}^{-gL}\obracket{\bar{\mathbb{Z}}_{2}}{\mathrm{e}^{i\hat{H}_{0}t}\mathrm{e}^{2g\hat{N}_{\uparrow}}\mathrm{e}^{-i\hat{H}_{0}t}}{\bar{\mathbb{Z}}_{2}}\nonumber \\
                                                                                  & =\sum_{m}\mathrm{e}^{-g\left(L-2N_{\uparrow}^{m}\right)}\left|\obracket m{\mathrm{e}^{-i\hat{H}_{0}t}}{\bar{\mathbb{Z}}_{2}}\right|^{2},
\end{align}
where $\obracket m{\mathrm{e}^{-i\hat{H}_{0}t}}{\bar{\mathbb{Z}}_{2}}$
is the unitary evolution using the PXP Hamiltonian. Since the number of up spins in the constrained subspace is at most $L/2$, for $g>0$ every weight satisfies $\mathrm{e}^{-g\left(L-2N_{\uparrow}^{m}\right)}\le1$, and therefore \emph{exactly}
\begin{align}
  \braket{\bar{\mathbb{Z}}_{2}\left(t\right)}{\bar{\mathbb{Z}}_{2}\left(t\right)}                                                                                                                                                      & \le1,\nonumber                                                                                                                \\
  P_{\bar{\mathbb{Z}}{}_{2}}\left(t\right)=\frac{\left|\obracket{\bar{\mathbb{Z}}_{2}}{\mathrm{e}^{-i\hat{H}_{0}t}}{\bar{\mathbb{Z}}_{2}}\right|^{2}}{\braket{\bar{\mathbb{Z}}_{2}\left(t\right)}{\bar{\mathbb{Z}}_{2}\left(t\right)}} & \ge\left|\obracket{\bar{\mathbb{Z}}_{2}}{\mathrm{e}^{-i\hat{H}_{0}t}}{\bar{\mathbb{Z}}_{2}}\right|^{2},\label{eq:enhancement}
\end{align}
where we used Eq.~\eqref{eq:generic_basis_state}. Namely, at each time $t$ the PXP return probability is enhanced by the  inverse of the smaller-than-unity normalization, and the same holds for $P_{\mathbb{Z}_{2}}(t)$. The only states with $N_{\uparrow}=L/2$ are the two N\'{e}el states, while the weight of states with $N_{\uparrow}=L/2-k$ is suppressed by $\mathrm{e}^{-2gk}$. Therefore, for $g\gg1$, the normalization is approximately
\begin{align}
  \braket{\bar{\mathbb{Z}}_{2}\left(t\right)}{\bar{\mathbb{Z}}_{2}\left(t\right)} & \approx\left|\obracket{\bar{\mathbb{Z}}_{2}}{\mathrm{e}^{-i\hat{H}_{0}t}}{\bar{\mathbb{Z}}_{2}}\right|^{2}\nonumber \\
                                                                                  & +\left|\obracket{\mathbb{Z}_{2}}{\mathrm{e}^{-i\hat{H}_{0}t}}{\bar{\mathbb{Z}}_{2}}\right|^{2},
\end{align}
such that $P_{\mathbb{Z}_{2}}(t)+P_{\bar{\mathbb{Z}}_{2}}(t)\approx1$: the wave packet is essentially composed only of the two N\'{e}el states, and the entire weight moves from one N\'{e}el state to the other, which makes the oscillations in Fig.~\ref{fig:probability-z2-z2bar} sharp and long-lived. This approximation holds as long as the PXP weight on the N\'{e}el states is large compared to the $\mathrm{e}^{-2g}$-suppressed weight on the other states, which is best satisfied at the revival times. At the first revival, the weight that PXP leaks out of the N\'{e}el states is dominated by states with $N_{\uparrow}=L/2-1$, and the infidelity is suppressed accordingly; indeed, we find numerically for $L=16,\ldots,\Lmax$ that $1-P_{\bar{\mathbb{Z}}_{2}}(t_{1})=c(g)\,L\,\mathrm{e}^{-2g}$, with $c(g)$ independent of $L$ for $g\gtrsim1$, valid as long as $c(g)L\mathrm{e}^{-2g}\ll1$, such that for $g=1$ the infidelity per site, $c(1)\mathrm{e}^{-2}\approx3.4\times10^{-4}$, is more than thirty times smaller than for PXP (Fig.~S1 of~\cite{SM}). At fixed $g$ the revivals thus still degrade with system size, and maintaining a fixed revival quality requires $g\gtrsim\frac{1}{2}\ln L$. Within the time window of Fig.~\ref{fig:probability-z2-z2bar}, however, the later revival peaks show no systematic dependence on $L$ up to $L=\Lmax$, remaining above $0.88$ for $g=1$ up to $t\approx95$, while for PXP they drop to $\approx0.1$--$0.2$ for $L\ge26$~\cite{SM}. At late times the PXP state thermalizes, and $P_{\bar{\mathbb{Z}}_{2}}$ approaches its value for random states in the same momentum sectors, which decreases exponentially with $L$ for any finite $g$~\cite{SM}. We note that for $g\gg1$ the ratio $P_{\bar{\mathbb{Z}}_{2}}\approx|a|^{2}/(|a|^{2}+|b|^{2})$, where $a=\obracket{\bar{\mathbb{Z}}_{2}}{\mathrm{e}^{-i\hat{H}_{0}t}}{\bar{\mathbb{Z}}_{2}}$ and $b=\obracket{\mathbb{Z}_{2}}{\mathrm{e}^{-i\hat{H}_{0}t}}{\bar{\mathbb{Z}}_{2}}$, is of order one for \emph{any} state, scarred or not, such that the revival heights are meaningful only compared to this random-state reference; for $g=1$ it decreases from $0.23$ for $L=16$ to $0.09$ for $L=26$, well below the observed revival peaks. For $g<0$, the weights grow with decreasing $N_{\uparrow}$, and the norm is dominated by states with few up spins; in particular, the all-down state gives $\braket{\bar{\mathbb{Z}}_{2}\left(t\right)}{\bar{\mathbb{Z}}_{2}\left(t\right)}\ge\mathrm{e}^{|g|L}\left|\obracket 0{\mathrm{e}^{-i\hat{H}_{0}t}}{\bar{\mathbb{Z}}_{2}}\right|^{2}$. The norm thus grows rapidly in time, and the return probabilities are suppressed by a factor that is exponentially small in $L$, as seen in Fig.~\ref{fig:probability-z2-z2bar}.

\emph{Physical interpretation and implementation.} Equations~\eqref{eq:generic_basis_state} and \eqref{eq:enhancement} show that the enhancement is a magnetization-resolved reweighting of the \emph{unitary} PXP dynamics: Any quantity diagonal in the computational basis, including $P_{\bar{\mathbb{Z}}_{2}}(t)$, follows from reweighting the snapshots of a Hermitian PXP (e.g., Rydberg) experiment~\cite{Bernien2017} in a way that for $g>0$ favors (suppresses the least) $\mathbb{Z}_2$ and $\bar{\mathbb{Z}}_2$. Renormalizing then simply multiplies all Fock states by the same factor so does not change their importance.

We now turn to physical implementation of the model. Rewriting $\hat{H}_{g}=\cosh(g)\hat{H}_{0}+i\sinh(g)\sum_{i}\hat{P}_{i-1}\hat{\sigma}_{i}^{y}\hat{P}_{i+1}$, we have that $\hat{H}_{g}-iL|\sinh g|$ is the no-click limit of a Lindblad evolution with the local, constrained jump operators $\hat{L}_{i}=\sqrt{2|\sinh g|}(1-s\hat{h}_{i})^{1/2}$, where $\hat{h}_{i}\equiv\hat{P}_{i-1}\hat{\sigma}_{i}^{y}\hat{P}_{i+1}$ and $s=\mathrm{sgn\left(g\right)}$. The constant shift drops out of all normalized quantities and is common to all postselected models, not just ours. Therefore our non-Hermitian Hamiltonian can be obtained by postselecting trajectories from a dissipative evolution (see Sec.~S1 of the Supplemental Material~\cite{SM}). Alternatively, by Eq.~\eqref{eq:similarity}, $\mathrm{e}^{-i\hat{H}_{g}t}\ket{\bar{\mathbb{Z}}_{2}}=\hat{K}_{g}\mathrm{e}^{-i\hat{H}_{0}t}\ket{\bar{\mathbb{Z}}_{2}}$ with $\hat{K}_{g}=\mathrm{e}^{g(\hat{N}_{\uparrow}-L/2)}$; for $g>0$, $\Vert\hat{K}_{g}\Vert\le1$, so a single filtering measurement at the final time postselects the non-Hermitian evolution from the Hermitian PXP dynamics, with success probability $\braket{\bar{\mathbb{Z}}_{2}(t)}{\bar{\mathbb{Z}}_{2}(t)}$.

\begin{figure}[t]
  \includegraphics[width=1\columnwidth]{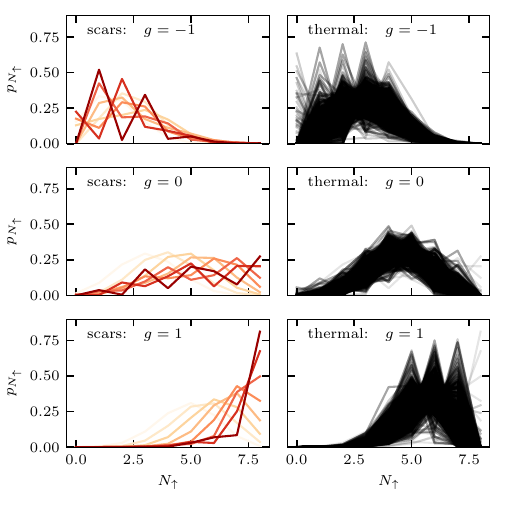}

  \caption{\label{fig:p_nup}Probability distribution $p_{N_{\uparrow}}$ to
    measure $N_{\uparrow}$ spins up in quantum scars (left column) and
    thermal states (right column) computed for $L=16$ and $g=-1$ (top
    row), $g=0$ (middle row) and $g=1$ (bottom row). In the left column,
    only scars with negative energies are plotted; darker colors correspond
    to eigenvalues closer to zero.}
\end{figure}

\emph{Non-Hermitian quantum scars}. The PXP model is known to have
a set of $L+1$ special eigenstates with atypically low entanglement
entropy, and atypically high overlap with the N\'{e}el states. In what follows, we show that the non-Hermitian model we consider enhances these features. Eq.~\eqref{eq:similarity} gives a relation between the right eigenvectors of $\hat{H}_{g}$ and the eigenvectors of $\hat{H}_{0}$,
\begin{equation}
  \hat{H}_{g}\left(\hat{V}_{\uparrow}\ket{\alpha}\right)=E_{\alpha}\left(\hat{V}_{\uparrow}\ket{\alpha}\right),
\end{equation}
where $\ket{\alpha}$ and $E_{\alpha}$ are the eigenvector-eigenvalue
pairs of $\hat{H}_{0}$. The similarity transformation Eq.~\eqref{eq:similarity_transformation}
redistributes the weight of components of the PXP eigenvectors according to the number of up spins. Defining $\hat{P}_{N_{\uparrow}}$ to be a projector on the subspace with $N_{\uparrow}$ up spins, the normalized weight of the right-eigenvector, $\hat{V}_{\uparrow}\ket{\alpha}$ in this subspace is,
\begin{equation}
  p_{N_{\uparrow}}\left(\alpha\right)=\frac{1}{Z}\obracket{\alpha}{\hat{V}_{\uparrow}^{\dag}\hat{P}_{N_{\uparrow}}\hat{V}_{\uparrow}}{\alpha}=\frac{\mathrm{e}^{2gN_{\uparrow}}}{Z}p_{N_{\uparrow}}^{\left(0\right)}\left(\alpha\right),\label{eq:up-spin-weight-distribution}
\end{equation}
where $Z=\sum_{N_{\uparrow}=0}^{L/2}\mathrm{e}^{2gN_{\uparrow}}p_{N_{\uparrow}}^{\left(0\right)}\left(\alpha\right)$
is the normalization factor, and $p_{N_{\uparrow}}^{\left(0\right)}\left(\alpha\right)=\obracket{\alpha}{\hat{P}_{N_{\uparrow}}}{\alpha}$,
is the weight of the PXP eigenvector in the corresponding subspace.
In the middle row of Fig.~\ref{fig:p_nup} we show the probability
distribution $p_{N_{\uparrow}}^{\left(0\right)}\left(\alpha\right)$
for all eigenvectors of a PXP model. The probability distribution
is divided into two \emph{qualitatively} different groups of states.
Quantum scar eigenvectors closer to the center of the spectrum show
a maximum probability for $N_{\uparrow}=L/2$, with probability decaying
for smaller values of $N_{\uparrow}$, while the probability distribution
for non-quantum scar states is peaked at around $N_{\uparrow}\approx0.28L$, the infinite-temperature value $(5-\sqrt{5})L/10$ for the constrained chain.
Interestingly, the further away are quantum scars from the center of the
spectrum the more similar they are to thermal states. The distributions
corresponding to non-Hermitian right eigenvectors are weighted by
$\frac{1}{Z}\mathrm{e}^{2gN_{\uparrow}}$ according to Eq.~\eqref{eq:up-spin-weight-distribution}.
For $g>0$ the weight of all probability distributions are compressed
towards $N_{\uparrow}=L/2$ (see bottom row of Fig.~\ref{fig:p_nup}),
while for $g<0$ they are compressed towards $N_{\uparrow}=0$ (see
top row of Fig.~\ref{fig:p_nup}). We stress that this reweighting concerns the right eigenvectors, normalized by themselves; biorthogonal quantities, such as the overlaps $\braket{\alpha_{L}}{\mathbb{Z}_{2}}\braket{\mathbb{Z}_{2}}{\alpha_{R}}$ and the biorthogonal entanglement entropy considered in Ref.~\cite{Chen2023WeakErgodicityBreaking}, are exactly independent of $g$, since $\hat{V}_{\uparrow}$ is diagonal and, for the entanglement, a product of single-site operators~\cite{SM}.

\begin{figure}[t]
  \includegraphics{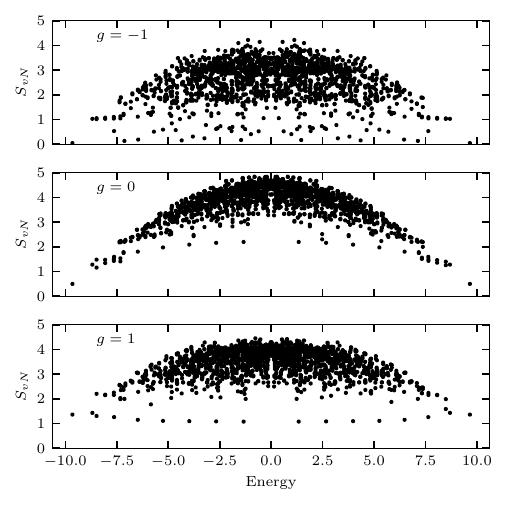}

  \caption{\label{fig:ee}Bipartite entanglement entropy (in bits) of all the right eigenvectors
    for $L=16$ and $g=-1,0,1$. For the exponentially degenerate $E=0$ manifold, the entanglement depends on the choice of basis within the manifold.}

\end{figure}
The bipartite entanglement is another metric used to distinguish between
scarred and non-scarred states. In Fig.~\ref{fig:ee} we show the
entanglement entropy of all right-eigenvectors for $g=-1,0,1$. Interestingly,
while in the PXP model ($g=0)$ the entanglement of the scars is locally
lower, but otherwise has an energy dependence, for moderate $g>0$ the entanglement entropy of \emph{all} the scars is close to 1\footnote{We choose to work with a logarithm in base 2 in the definition of the
  entanglement entropy.}: for $L=16$ and $g=1$, the $L$ scars with nonzero energy have an entanglement entropy between $1.08$ and $1.44$, while all other nonzero-energy states, except two near the spectral edges, have larger entanglement, with a median of $3.6$. This can be used to distinguish the scars from the thermal
states. For $g<0$ there is a group of states with close to zero entanglement
entropy, and a group of states with entanglement entropy close to 1, while the remaining states have a large entanglement entropy.

These features can be understood from the limits $g\to\pm\infty$, in which every right eigenvector collapses onto its component with the largest (smallest) number of up spins~\cite{SM}. For $g\to\infty$, every eigenvector with a nonzero overlap with the N\'{e}el states, scarred or not, approaches a symmetric or antisymmetric superposition of the two N\'{e}el states, with an entanglement entropy of exactly 1, while the remaining eigenvectors approach their components with the largest number of up spins; the scars, which have the largest N\'{e}el weight, approach this limit first, which is the origin of the separation at intermediate $g$. For larger $g$ this separation degrades, as all eigenvectors in the same symmetry sector become nearly parallel. For $g\to-\infty$, eigenvectors approach either the all-down state, with zero entanglement, or single-magnon states with a definite quasi-momentum, with an entanglement entropy of exactly 1~\cite{SM}.

\emph{Conclusions}. We have shown that biasing the spin flips in the PXP model enhances the revivals of the N\'{e}el states, and that this enhancement can be understood exactly. Although the spectrum and eigenvectors of PXP are not known in closed form, the mapping expresses the enhanced revivals and eigenvector weights directly in terms of PXP quantities, and pinpoints the mechanism: a magnetization-resolved reweighting of the \emph{unchanged} PXP dynamics, which concentrates the weight on the N\'{e}el states and suppresses the first-revival infidelity, asymptotically by a factor $\propto\mathrm{e}^{-2g}$. This explains the numerical observations of Ref.~\cite{Chen2023WeakErgodicityBreaking}, provides an exact framework for those of Ref.~\cite{Shen2024EnhancedManyBody}, and in particular shows that the divergence of the revival period near the exceptional point is an overall rescaling of energy. The mapping applies whenever the Hamiltonian can be written as $\mathrm{e}^{g}\hat{H}_{+}+\mathrm{e}^{-g}\hat{H}_{+}^{\dagger}+\hat{H}_{d}$, with Hermitian $\hat{H}_{d}$, $[\hat{N},\hat{H}_{+}]=\hat{H}_{+}$, and $[\hat{N},\hat{H}_{d}]=0$ for some Hermitian operator $\hat{N}$; an enhancement of revivals further requires $\hat{N}$ to be bounded, with its maximum attained on the states of interest, as for $N_{\uparrow}$ and the N\'{e}el states here. It therefore applies to any kinetically constrained model or LGT satisfying these conditions. Perturbations that break the mapping spoil its exactness, but the enhancement persists for weak perturbations, which already damp the PXP revivals considerably, and is lost for strong ones (Table~S1 of~\cite{SM}).

The same exactness also clarifies the limitations of the effect: it concerns normalized, postselected quantities, and its realization as a no-click trajectory is exponentially costly in time and system size, while the unconditional dissipative dynamics shows no enhancement (Sec.~S1 of~\cite{SM}). Its most direct realization is therefore the final-time filtering, or reweighting of snapshots, of Hermitian PXP dynamics (Sec.~S1 of~\cite{SM}). Whether a related enhancement can be obtained without postselection, as for periodic projective measurements~\cite{Paviglianiti2025EnhancingRevivals}, is an interesting question for future work.

\medskip
\footnotesize
\emph{Acknowledgments}. The authors thank Zlatko Papić for fruitful discussions. Y.B.L.~acknowledges support by the Israel Science Foundation (grant No.~1304/23). J.C.H.~acknowledges funding by the Max Planck Society, the Deutsche Forschungsgemeinschaft (DFG, German Research Foundation) under Germany's Excellence Strategy -- EXC-2111 -- 390814868, and the European Research Council (ERC) under the European Union's Horizon Europe research and innovation program (Grant Agreement No.~101165667)---ERC Starting Grant QuSiGauge. Views and opinions expressed are, however, those of the author(s) only and do not necessarily reflect those of the European Union or the European Research Council Executive Agency. Neither the European Union nor the granting authority can be held responsible for them. This work is part of the Quantum Computing for High-Energy Physics (QC4HEP) working group. A.L.~acknowledges support from the Leverhulme Trust Research Project Grant RPG-2025-063.

\bibliography{biblio,references}

\clearpage
\beginsupplement
\onecolumngrid
\begin{center}
  \textbf{\large Supplemental Material for ``Non-Hermiticity-Enhanced Quantum Revivals in a Lattice Gauge Theory''}
\end{center}

\section*{S1. Exact no-click representation, its cost, the unconditional dynamics, and the final-time filter}

Using $\hat{\sigma}^{\pm}=(\hat{\sigma}^{x}\pm i\hat{\sigma}^{y})/2$ and $e^{\pm g}=\cosh g\pm\sinh g$,
\begin{equation}
  \hat{H}_{g}=\cosh(g)\,\hat{H}_{0}+i\sinh(g)\sum_{i=1}^{L}\hat{h}_{i},\qquad\hat{h}_{i}\equiv\hat{P}_{i-1}\hat{\sigma}_{i}^{y}\hat{P}_{i+1}.\label{eq:decomposition}
\end{equation}
Since $\hat{P}_{i\pm1}$ commute with $\hat{\sigma}_{i}^{y}$ and $(\hat{\sigma}_{i}^{y})^{2}=1$, $\hat{h}_{i}^{2}=\hat{Q}_{i}\equiv\hat{P}_{i-1}\hat{P}_{i+1}$, a projector. Hence $\hat{h}_{i}$ has eigenvalues $0,\pm1$, so it describes both gain and loss. With $s=\mathrm{sgn}(g)$, the operator $1-s\hat{h}_{i}$ is positive semidefinite, and its square root is the local three-site operator
\begin{equation}
  (1-s\hat{h}_{i})^{1/2}=1-\frac{s}{\sqrt{2}}\hat{h}_{i}-\left(1-\frac{1}{\sqrt{2}}\right)\hat{Q}_{i}=(1-\hat{Q}_{i})+\sqrt{2}\,\hat{Q}_{i}\hat{\Pi}_{i}^{(-s)},
\end{equation}
where $\hat{\Pi}_{i}^{(-s)}$ projects site $i$ onto the eigenstate of $\hat{\sigma}_{i}^{y}$ with eigenvalue $-s$. Defining $\hat{L}_{i}=\sqrt{2|\sinh g|}\,(1-s\hat{h}_{i})^{1/2}$, we have $\hat{L}_{i}^{\dagger}\hat{L}_{i}=2|\sinh g|(1-s\hat{h}_{i})$, and therefore, exactly,
\begin{equation}
  \hat{K}\equiv\cosh(g)\hat{H}_{0}-\frac{i}{2}\sum_{i}\hat{L}_{i}^{\dagger}\hat{L}_{i}=\hat{H}_{g}-iL|\sinh g|.\label{eq:no-click}
\end{equation}
$\hat{K}$ is the no-click generator of the Lindblad equation $\dot{\hat{\rho}}=-i[\cosh(g)\hat{H}_{0},\hat{\rho}]+\sum_{i}\big(\hat{L}_{i}\hat{\rho}\hat{L}_{i}^{\dagger}-\frac{1}{2}\{\hat{L}_{i}^{\dagger}\hat{L}_{i},\hat{\rho}\}\big)$~\cite{daley_quantum_2014}. The idea of rendering the anti-Hermitian part dissipative by a constant imaginary shift was put forward in Ref.~\cite{Chen2023WeakErgodicityBreaking}, where the jump operators were obtained perturbatively in a large shift; here they are exact and local. Since $\mathrm{e}^{-i\hat{H}_{g}t}=\mathrm{e}^{L|\sinh g|t}\mathrm{e}^{-i\hat{K}t}$, the shift cancels in all normalized quantities. We verified Eqs.~\eqref{eq:similarity}, \eqref{eq:decomposition}, and \eqref{eq:no-click} numerically to machine precision.

\emph{Cost}. The probability of no jump up to time $t$, starting from $\ket{\bar{\mathbb{Z}}_{2}}$, is $p_{\mathrm{nc}}(t)=\Vert\mathrm{e}^{-i\hat{K}t}\ket{\bar{\mathbb{Z}}_{2}}\Vert^{2}=\mathrm{e}^{-2L|\sinh g|t}\braket{\bar{\mathbb{Z}}_{2}(t)}{\bar{\mathbb{Z}}_{2}(t)}$, e.g., $p_{\mathrm{nc}}\approx\mathrm{e}^{-199}$ after one revival period, $T\approx4.7$, for $L=18$ and $g=1$. The extensive exponent cannot be avoided by another choice of jump operators: any representation $\hat{H}_{g}=\hat{H}'-\frac{i}{2}\sum_{j}\hat{L}_{j}^{\prime\dagger}\hat{L}'_{j}+ic$ with Hermitian $\hat{H}'$ requires $c\ge|\sinh g|\,\lambda_{\max}$, where $\lambda_{\max}$ is the largest eigenvalue of $s\sum_{i}\hat{h}_{i}$. Since $\sum_{i}\hat{h}_{i}$ is mapped onto $\hat{H}_{0}$ by the rotation $\mathrm{e}^{-i\pi\sum_{i}\hat{\sigma}_{i}^{z}/4}$, which commutes with the projectors, $\lambda_{\max}$ is the largest eigenvalue of PXP, $\lambda_{\max}\approx0.60L$, such that $p_{\mathrm{nc}}\lesssim\mathrm{e}^{-120}$ after one period for $L=18$ and $g=1$.

\emph{Unconditional dynamics}. Since the $\hat{L}_{i}$ are Hermitian, the Lindbladian is unital and relaxes to the infinite-temperature state of the constrained subspace. Integrating it for $L=10$ and $g=1$, starting from $\ket{\bar{\mathbb{Z}}_{2}}$, we find that $\langle\bar{\mathbb{Z}}_{2}|\hat{\rho}(t)|\bar{\mathbb{Z}}_{2}\rangle$ never exceeds $0.011$ for $t>2$, close to its infinite-temperature value $1/\mathcal{D}=0.008$, where $\mathcal{D}=123$ is the dimension of the constrained subspace: the unconditional dynamics shows no revivals. In contrast, for jump operators of the form $\hat{L}=a+b\hat{h}_{i}$ with a large constant shift, as in Ref.~\cite{Chen2023WeakErgodicityBreaking}, the dissipator reduces to $b^{2}\mathcal{D}[\hat{h}_{i}]$ with $b^{2}\to0$, and the unconditional dynamics approaches weakly dephased PXP dynamics, whose revivals are those of PXP, not enhanced ones.

\emph{Final-time filter}. By Eq.~\eqref{eq:similarity}, $\mathrm{e}^{-i\hat{H}_{g}t}\ket{\bar{\mathbb{Z}}_{2}}=\hat{K}_{g}\mathrm{e}^{-i\hat{H}_{0}t}\ket{\bar{\mathbb{Z}}_{2}}$ with $\hat{K}_{g}=\mathrm{e}^{g(\hat{N}_{\uparrow}-L/2)}$. For $g>0$, $\Vert\hat{K}_{g}\Vert\le1$, such that $\hat{K}_{g}$ is the Kraus operator of a heralded (filtering) measurement applied once, at the final time, to the Hermitian PXP state. Its success probability is $\Vert\hat{K}_{g}\mathrm{e}^{-i\hat{H}_{0}t}\ket{\bar{\mathbb{Z}}_{2}}\Vert^{2}=\braket{\bar{\mathbb{Z}}_{2}(t)}{\bar{\mathbb{Z}}_{2}(t)}$, which is of order one near the revivals (e.g., $0.87$ at the first revival for $L=10$, $g=1$), but exponentially small in $L$ at generic times.

\section*{S2. Relation to the models of Refs.~\cite{Chen2023WeakErgodicityBreaking} and \cite{Shen2024EnhancedManyBody}}

The non-Hermitian PXP model of Ref.~\cite{Chen2023WeakErgodicityBreaking} reads $\hat{H}_{\gamma}+h\sum_{j}\hat{\sigma}_{j}^{z}$ with $\hat{H}_{\gamma}=\sum_{j}\hat{P}_{j-1}(\hat{\sigma}_{j}^{x}+i\gamma\hat{\sigma}_{j}^{y})\hat{P}_{j+1}=\sum_{j}\hat{P}_{j-1}[(1+\gamma)\hat{\sigma}_{j}^{+}+(1-\gamma)\hat{\sigma}_{j}^{-}]\hat{P}_{j+1}$. Matching the ratio of the raising and lowering amplitudes, $\mathrm{e}^{2g}=(1+\gamma)/(1-\gamma)$, gives $\gamma=\tanh g$ and $\hat{H}_{\gamma}=\sqrt{1-\gamma^{2}}\,\hat{H}_{g}$. Since the longitudinal field is diagonal in the computational basis, it commutes with $\hat{V}_{\uparrow}$, and
\begin{equation}
  \hat{H}_{\gamma}+h\sum_{j}\hat{\sigma}_{j}^{z}=\sqrt{1-\gamma^{2}}\;\hat{V}_{\uparrow}\Big(\hat{H}_{0}+h_{\mathrm{eff}}\sum_{j}\hat{\sigma}_{j}^{z}\Big)\hat{V}_{\uparrow}^{-1},\qquad h_{\mathrm{eff}}=\frac{h}{\sqrt{1-\gamma^{2}}}.
\end{equation}
For $|\gamma|<1$ the model is thus similar to the Hermitian PXP model in a rescaled field: its spectrum is real, the revival period is $T=T_{\mathrm{PXP}}/\sqrt{1-\gamma^{2}}$, which diverges as $\gamma\to1$, and its phase boundaries, as well as the fragility of the scars to $h$, are those of the Hermitian model at $h_{\mathrm{eff}}$, which grows as $\gamma\to1$. In units of the hopping of $\hat{H}_{g}$, the period is independent of $g$. For $|\gamma|>1$, the same construction holds with a complex $g$ (with imaginary part $\pm\pi/2$, depending on the branch), such that $\hat{V}_{\uparrow}$ is no longer Hermitian, the prefactor becomes $\pm i\sqrt{\gamma^{2}-1}$, and $h_{\mathrm{eff}}$ is imaginary; at $h=0$ the spectrum is purely imaginary. Since $\hat{V}_{\uparrow}=\prod_{j}\mathrm{e}^{g\hat{n}_{j}}$ is a product of single-site operators, the biorthogonal reduced density matrix of a right/left eigenvector pair satisfies $\mathrm{Tr}_{B}\ket{\alpha_{R}}\bra{\alpha_{L}}=\hat{V}_{A}\hat{\rho}_{A}^{(0)}\hat{V}_{A}^{-1}$, where $\hat{\rho}_{A}^{(0)}$ is the PXP reduced density matrix, such that the biorthogonal entanglement entropy is exactly that of PXP for all $g$.

The models of Ref.~\cite{Shen2024EnhancedManyBody} are mapped onto Hermitian models in the same way. Their model with non-Hermiticity $\hat{\sigma}^{x}+iu\hat{\sigma}^{y}$ on even sites only is similar, via $\mathrm{e}^{g\hat{N}_{\uparrow}^{\mathrm{even}}}$ with $\tanh g=u$, to a Hermitian PXP chain with amplitude $\sqrt{1-u^{2}}$ on even sites and $1$ on odd sites, while their model with $\hat{\sigma}^{x}+iu\hat{\sigma}^{y}$ on even and $\hat{\sigma}^{x}-iu\hat{\sigma}^{y}$ on odd sites is similar to $\sqrt{1-u^{2}}\,\hat{H}_{0}$ via $\mathrm{e}^{g(\hat{N}_{\uparrow}^{\mathrm{even}}-\hat{N}_{\uparrow}^{\mathrm{odd}})}$. Their translation-invariant model is ours. All these statements were verified numerically to machine precision.

\section*{S3. Finite-size analysis}

\begin{figure}[h]
  \includegraphics[width=\textwidth]{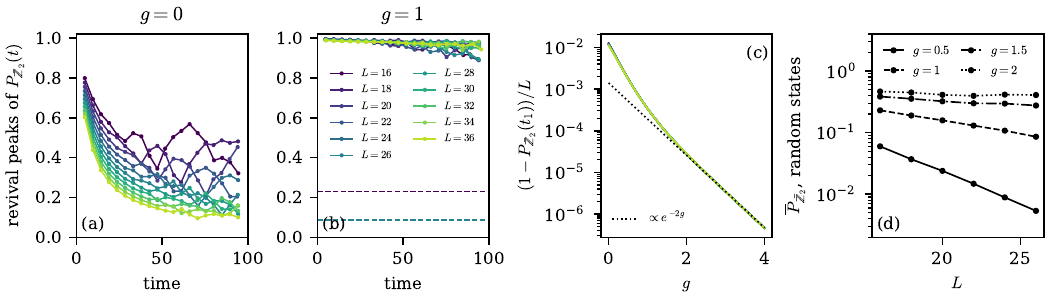}
  \caption{\label{fig:fss}(a,b) Heights of the revival peaks of $P_{\bar{\mathbb{Z}}_{2}}(t)$ for $g=0$ and $g=1$ and $L=16,\ldots,\Lmax$ (legend in panel (b) applies to all panels), with initial state $\ket{\bar{\mathbb{Z}}_{2}}$. Dashed lines in (b): average of $P_{\bar{\mathbb{Z}}_{2}}$ over random states in the $k=0,\pi$ sectors, $\overline{P}_{\bar{\mathbb{Z}}_{2}}$, for $L=16$ and $L=26$. (c) First-revival infidelity per site as a function of $g$, for all $L$; the curves collapse, and approach $\propto\mathrm{e}^{-2g}$ (dotted). (d) $\overline{P}_{\bar{\mathbb{Z}}_{2}}$ as a function of $L$.}
\end{figure}

By Eq.~\eqref{eq:generic_basis_state} and the expression for the norm in the main text, $P_{\mathbb{Z}_{2}}(t)$ and $P_{\bar{\mathbb{Z}}_{2}}(t)$ for \emph{any} $g$ follow from a single Hermitian PXP evolution, through the two N\'{e}el amplitudes and the weights $w_{N_{\uparrow}}(t)=\sum_{m:N_{\uparrow}^{m}=N_{\uparrow}}|\obracket m{\mathrm{e}^{-i\hat{H}_{0}t}}{\bar{\mathbb{Z}}_{2}}|^{2}$, namely $\braket{\bar{\mathbb{Z}}_{2}(t)}{\bar{\mathbb{Z}}_{2}(t)}=\sum_{N_{\uparrow}}\mathrm{e}^{-g(L-2N_{\uparrow})}w_{N_{\uparrow}}(t)$. Since $\ket{\bar{\mathbb{Z}}_{2}}$ is an equal superposition of states with quasi-momenta $0$ and $\pi$, we perform the evolution in these two sectors only, which allows us to reach $L=\Lmax$ (sector dimensions of about $9.3\times10^{5}$). We verified that this reproduces Fig.~\ref{fig:probability-z2-z2bar} for $L=18$, and that it agrees with a direct evolution in the full Hilbert space to machine precision for smaller sizes. The revival times are obtained from a linear fit to the first six PXP revivals, giving a period $T=4.715$ for all $L\ge22$, and the peak heights are the maxima of $P_{\bar{\mathbb{Z}}_{2}}$ within $\pm T/4$ of each revival time.

Figure~\ref{fig:fss}(c) shows that the infidelity at the first revival obeys $1-P_{\bar{\mathbb{Z}}_{2}}(t_{1})=c(g)\,L\,\mathrm{e}^{-2g}$, with $c(g)$ independent of $L$ to three significant digits for $g\gtrsim1$ (e.g., $c(1)=2.49\times10^{-3}$ for all $L$) (for $g=0$ it drifts by $\approx12\%$ between $L=16$ and $L=\Lmax$), and $c(g)\approx1.4\times10^{-3}$ for $g\gtrsim3$, consistent with leading-order leakage into states with $N_{\uparrow}=L/2-1$, whose weight is suppressed by $\mathrm{e}^{-2g}$, i.e., $1-P_{\bar{\mathbb{Z}}_{2}}(t_{1})\approx\mathrm{e}^{-2g}w_{L/2-1}(t_{1})/(|a|^{2}+|b|^{2})$, which is valid as long as this quantity is small. For PXP the infidelity per site is $\approx0.012$, while for $g=1$ it is $c(1)\mathrm{e}^{-2}\approx3.4\times10^{-4}$. Figures~\ref{fig:fss}(a,b) show that, for PXP, the later revival peaks decay faster with increasing $L$, reaching $\approx0.1$--$0.2$ at $t\approx95$ for $L\ge26$, whereas for $g=1$ they show no systematic dependence on $L$ within this time window, and remain above $0.88$ for all $L$. These peaks are well above the random-state reference, which for $g=1$ decreases from $0.23$ for $L=16$ to $0.09$ for $L=26$; we computed this reference, which requires sampling many random states in the full symmetry sectors, only up to $L=26$, where its exponential decay is already evident. Figure~\ref{fig:fss}(d) shows that this reference decreases exponentially with $L$ for any finite $g$, but increases with $g$, approaching $1/2$ for $g\to\infty$, when $P_{\bar{\mathbb{Z}}_{2}}\approx|a|^{2}/(|a|^{2}+|b|^{2})$ for any state. The enhancement is therefore meaningful for moderate $g$, and, at fixed $g$, the revivals ultimately decay in the thermodynamic limit, albeit much more slowly than for PXP.

\section*{S4. Entanglement of the right eigenvectors for $|g|\to\infty$}

As $g\to\pm\infty$, each right eigenvector $\hat{V}_{\uparrow}\ket{\alpha}$ is dominated by its component with the largest (smallest) $N_{\uparrow}$. For $g\to\infty$, every eigenvector with a nonzero N\'{e}el overlap approaches $(\ket{\mathbb{Z}_{2}}\pm\ket{\bar{\mathbb{Z}}_{2}})/\sqrt{2}$, the sign being fixed by its quasi-momentum, whose entanglement entropy is exactly 1. This holds for scarred and thermal eigenvectors alike, as the right eigenvectors become nearly parallel; the condition number of $\hat{V}_{\uparrow}$ is $\mathrm{e}^{|g|L/2}$. For $g\to-\infty$, eigenvectors with a nonzero overlap with the all-down state, which lie in the zero-momentum, inversion-symmetric sector, converge to it and have zero entanglement. Most other eigenvectors converge to single-magnon states $\sum_{j}\mathrm{e}^{ikj}\ket{j}/\sqrt{L}$ with one up spin, whose half-chain entanglement entropy is exactly 1, while the few eigenvectors without weight on states with zero or one up spin retain a larger entanglement entropy. For $L=12$ and $g=-8$, we find entanglement entropies of $0$ for the $23$ eigenvectors of the first kind, between $0.98$ and $1.01$ for the $289$ eigenvectors of the second kind, and above $2.2$ for the remaining $10$.

\section*{S5. Robustness to perturbations that break the mapping}

To test the enhancement beyond the exact mapping, we add to $\hat{H}_{g}$ a Hermitian perturbation $\lambda\,\delta\hat{H}$, which is not of the form discussed in the Conclusions, such that $\hat{V}_{\uparrow}^{-1}(\hat{H}_{g}+\lambda\delta\hat{H})\hat{V}_{\uparrow}$ is no longer Hermitian. We consider a translation-invariant perturbation, $\delta\hat{H}_{1}=\sum_{i}\hat{P}_{i-1}\hat{\sigma}_{i}^{x}\hat{P}_{i+1}(\hat{Z}_{i-2}+\hat{Z}_{i+2})$, with $\lambda>0$, the sign opposite to the one that stabilizes the PXP revivals~\cite{Choi2019EmergentSU2}, and random Rabi frequencies, $\delta\hat{H}_{2}=\sum_{i}\epsilon_{i}\hat{P}_{i-1}\hat{\sigma}_{i}^{x}\hat{P}_{i+1}$ with $\epsilon_{i}$ uniform in $[-1,1]$, averaged over $10$ realizations, and evolve $\ket{\bar{\mathbb{Z}}_{2}}$ directly in the full constrained subspace for $L=20$. Table~\ref{tab:robust} lists the heights of the first, fifth, and tenth revival peaks of $P_{\bar{\mathbb{Z}}_{2}}(t)$. For weak perturbations ($\lambda\lesssim0.05$), which already damp the PXP revivals considerably, the enhanced revivals for $g=1$ remain above $0.8$ for at least five revivals; for stronger perturbations the later revivals decay towards the random-state reference ($\approx0.16$ for $L=20$, $g=1$), such that the enhancement does not protect the revivals against strong perturbations. Results for $L=16$ are similar. In contrast, diagonal perturbations, such as the random on-site fields $\sum_{j}w_{j}\hat{Z}_{j}$ considered in Ref.~\cite{Shen2024EnhancedManyBody}, commute with $\hat{V}_{\uparrow}$ and preserve the mapping, so that the bound of Eq.~\eqref{eq:enhancement} guarantees an enhancement over the correspondingly perturbed Hermitian PXP model.

\begin{table}[h]
  \caption{\label{tab:robust}Revival peaks $n=1,5,10$ of $P_{\bar{\mathbb{Z}}_{2}}(t)$ for $\hat{H}_{g}+\lambda\delta\hat{H}$, $L=20$. For $\delta\hat{H}_{2}$, the mean over $10$ disorder realizations is given, with the standard deviation of the fifth and tenth peaks in parentheses.}
  \begin{ruledtabular}
    \begin{tabular}{lccc}
      perturbation        & $\lambda$ & $g=0$                      & $g=1$\tabularnewline
      \hline
      none                & $0$       & $0.76,\ 0.45,\ 0.28$       & $0.99,\ 0.99,\ 0.96$\tabularnewline
      $\delta\hat{H}_{1}$ & $0.02$    & $0.61,\ 0.34,\ 0.27$       & $0.98,\ 0.98,\ 0.97$\tabularnewline
      $\delta\hat{H}_{1}$ & $0.05$    & $0.45,\ 0.24,\ 0.04$       & $0.96,\ 0.95,\ 0.63$\tabularnewline
      $\delta\hat{H}_{1}$ & $0.1$     & $0.28,\ 0.12,\ 0.06$       & $0.89,\ 0.86,\ 0.58$\tabularnewline
      $\delta\hat{H}_{2}$ & $0.05$    & $0.66,\ 0.17(4),\ 0.04(1)$ & $0.97,\ 0.84(7),\ 0.55(7)$\tabularnewline
      $\delta\hat{H}_{2}$ & $0.1$     & $0.47,\ 0.02(1),\ 0.01(0)$ & $0.89,\ 0.35(16),\ 0.35(15)$\tabularnewline
      $\delta\hat{H}_{2}$ & $0.2$     & $0.13,\ 0.00(0),\ 0.00(0)$ & $0.39,\ 0.16(12),\ 0.14(13)$\tabularnewline
    \end{tabular}
  \end{ruledtabular}
\end{table}

\end{document}